 \documentclass[preprint,onecolumn,amsmath,amssymb,aps,prb]{revtex4-1}

\usepackage{graphicx}
\usepackage{dcolumn}
\usepackage{bm}
 \usepackage{float} 

\usepackage{fullpage}
\usepackage{color}
\usepackage{subcaption}
\usepackage{multirow}
\usepackage{amsmath}
\usepackage{amssymb}
\usepackage{amsthm}
\usepackage[colorlinks   = true, urlcolor     = cyan, linkcolor    = blue, citecolor   = red]{hyperref}
\usepackage{bm}
\usepackage{pict2e}
\usepackage{epstopdf}
\usepackage{comment}
\usepackage{epsfig}
\usepackage{url}
\usepackage[ruled,vlined]{algorithm2e}

\usepackage{cancel}
\usepackage{amsmath}
\usepackage{amssymb}
\usepackage{etoolbox}
\usepackage{graphicx}
 \usepackage[english]{babel}
\usepackage{mathrsfs} 
\usepackage{enumitem}                                  

\usepackage{amsthm} 
\usepackage{siunitx}
\usepackage{ upgreek }

\begin{document}


\title{Elastic properties of Janus transition metal dichalcogenide nanotubes from first principles}

\author{Arpit Bhardwaj}
\affiliation{College of Engineering, Georgia Institute of Technology, Atlanta, GA 30332, USA}

\author{Phanish Suryanarayana}
\email{phanish.suryanarayana@ce.gatech.edu}
\affiliation{College of Engineering, Georgia Institute of Technology, Atlanta, GA 30332, USA}


\begin{abstract}
We calculate the elastic properties of Janus transition metal dichalcogenide (TMD) nanotubes using first principles Kohn-Sham density functional theory (DFT). Specifically,  we perform electronic structure simulations that exploit the cyclic and helical symmetry in the system to compute the Young's moduli, Poisson's ratios, and torsional moduli for twenty-seven select armchair and zigzag  Janus TMD nanotubes at their equilibrium diameters. We find the following trend in the moduli values: MSSe $>$ MSTe $>$ MSeTe, while their anisotropy with respect to armchair and zigzag configurations has the following ordering:  MSTe $>$ MSeTe $>$ MSSe. This anisotropy and its ordering between the different groups is confirmed by computing the shear modulus from the torsional modulus using an isotropic elastic continuum model, and comparing it against the value predicted from the isotropic relation featuring the Young's modulus and Poisson's ratio. We also develop a model for the Young's and torsional moduli of Janus TMD nanotubes based on linear regression. 
\end{abstract}


\maketitle

\section{Introduction} Nanotubes are tubular structures with diameters that typically range from one to few tens of nanometers. They have been the subject of a number of research efforts, motivated by their enhanced/novel mechanical, electronic, thermal, and optical properties, especially when compared  to their bulk counterparts \cite{tenne2003advances, rao2003inorganic, serra2019overview}. In particular, since the landmark realization of carbon nanotubes in the year 1991 \cite{iijima1991helical}, more than thirty distinct nanotubes have  been synthesized, with many more likely in the future. This is because nanotubes can be constructed  from their two-dimensional counterparts (atleast in concept), thousands of which have  been predicted to be stable \cite{zhou20192dmatpedia, gjerding2021recent, haastrup2018computational} from first principles Kohn-Sham density functional theory (DFT) \cite{kohn1965self, hohenberg1964inhomogeneous} calculations. 

Transition metal dichalcogenide (TMD) nanotubes --- one-dimensional analogues of the two-dimensional materials of the form MX\textsubscript{2}, where  M and  X are used to represent the transition metal and chalcogen, respectively --- are currently the most varied group, with the highest number of materials synthesized to date \cite{tenne2003advances, rao2003inorganic, serra2019overview}. In particular, they are known for their large tensile strength \cite{kis2003shear, kaplan2007mechanical, kaplan2006mechanical, tang2013revealing, bhardwaj2021torsional} and electronic properties that can be tuned through mechanical deformations \cite{zibouche2014electromechanical, oshima2020geometrical, li2014strain, ghorbani2013electromechanics, lu2012strain,ansari2015ab, levi2015nanotube, 10.1088/1361-6528/ac1a90}. However, most of these nanotubes are multi-walled  and have large diameters, which limits the appearance of novel and exotic properties found in small diameter single-walled nanotubes susceptible to quantum confinement effects. Furthermore, only a small fraction of the total number of possible TMD nanotubes have been synthesized to date, which can be partly attributed to the relatively high bending energy associated with TMD nanotubes \cite{kumar2020bending}. 

The recent synthesis of Janus TMD monolayers \cite{lu2017janus, zhang2017janus, trivedi2020room, lin2020low} --- materials of the form MXY, where X and Y represent different chalcogens --- presents an opportunity to  overcome some of the aforementioned limitations of TMD nanotubes. In particular, the asymmetry introduced by having different chalcogens on either side of the monolayer makes the nanotube configuration more energetically favorable \cite{xiong2018spontaneous}, likely making them easier to synthesize, particularly given that there exists an energy minimizing  nanotube diameter. This has motivated a number of studies for predicting the electronic \cite{oshima2020geometrical, wang2018mechanical, tang2018janus, mikkelsen2021band, luo2019electronic, zhao2015ultra} and optical \cite{ju2021tuning, ju2021rolling, zhang2019mosse, tang2018janus, oshima2020geometrical, xie2021theoretical} properties of Janus TMD nanotubes from first principles DFT calculations. Janus TMD nanotubes are expected to have a number of  technological applications, including optoelectric \cite{yagmurcukardes2020quantum, tang2018janus, oshima2020geometrical, xie2021theoretical} and nanoelectromechanical (NEMS) devices \cite{yudilevichself, levi2015nanotube, divon2017torsional}, as well as nanocomposites reinforcement \cite{shtein2013fracture, otorgust2017important, simic2019impact, nadiv2016critical, huang2016advanced, naffakh2016polymer}. In all cases, an accurate characterization of the elastic properties is important for the design process. However, apart from Ref.~\cite{wang2018mechanical} where the Young's modulus of the MoSSe nanotube has been computed using DFT, the mechanical properties of Janus TMD nanotubes remain unexplored heretofore. 

In this work, we compute the elastic properties of single-walled Janus TMD nanotubes from first principles. Specifically, considering the zigzag and armchair versions of the twenty-seven Janus TMD nanotube that have previously been predicted to be stable, we calculate the Young's moduli, Poisson's ratios, and torsional moduli for these materials at their equilibrium diameters, all using cyclic and helical symmetry-adapted Kohn-Sham DFT. We find the following trend in the moduli: MSSe $>$ MSTe $>$ MSeTe, while their anisotropy with respect to armchair and zigzag configurations has the following ordering:  MSTe $>$ MSeTe $>$ MSSe. This anisotropy and its ordering between the different groups is confirmed by computing the shear modulus value from the torsional modulus using an isotropic elastic continuum model, and comparing it with that predicted by the isotropic relation featuring the Young's modulus and Poisson's ratio.  We also develop a model for the Young's and  torsional moduli of  the nanotubes based on linear regression, with the following features: metal-chalcogen bonds' nature/characteristics and the difference in electronegativity between the chalcogens.  

The remainder of this article is organized as follows. In Section~\ref{Sec:Methods}, we list the  Janus TMD nanotubes selected  and describe the calculation of their elastic properties using symmetry-adapted Kohn-Sham DFT simulations.  The results so obtained are presented and discussed  in Section~\ref{Sec:Results}. Finally, we conclude in Section~\ref{Sec:Conclusions}.


\section{Systems and methods} \label{Sec:Methods}

We consider twenty-seven materials that represents the set of all  single-walled Janus TMD nanotubes that have previously been predicted to be stable \cite{bolle2021structural}. Specifically, we consider the following nanotubes: (i) M$=$\{V, Nb, Ta, Cr, Mo, W\} and X,Y$=$\{S, Se, Te\} with 2H-t symmetry \cite{nath2002nanotubes, bandura2014tis2}, and (ii) M$=$\{Ti, Zr, Hf\} and X,Y$=$\{S, Se, Te\} with 1T-o symmetry \cite{nath2002nanotubes, bandura2014tis2}, all in both armchair and zigzag configurations, with the heavier chalcogen on the outside.  The diameters for these nanotubes are selected so as to minimize the ground state Kohn-Sham energy \cite{bolle2021structural}, since experimentally synthesized nanotubes are likely to adopt energy minimizing configurations. 

\begin{figure}[htbp!]
\centering
\includegraphics[width=0.5\textwidth]{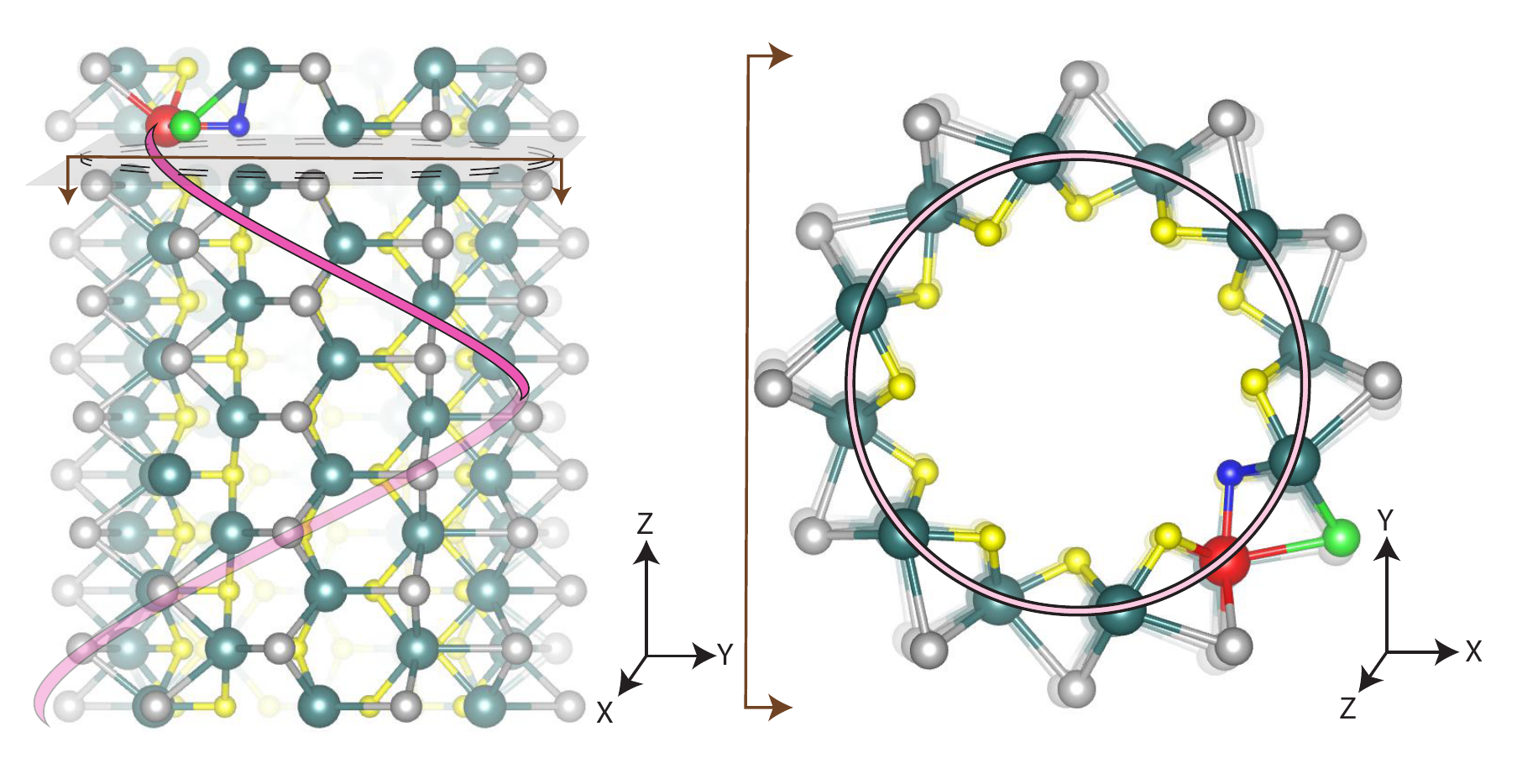}
\caption{Illustration of a twisted (6,6) TMD Janus nanotube with 2H-t symmetry, generated using VESTA \cite{momma2008vesta}. In particular, every atom in the nanotube is a cyclic and/or helical image of one of the three fundamental domain atoms: the metal and chalcogen atoms colored red and blue/green, respectively. This cyclic and helical symmetry is exploited while performing Kohn-Sham simulations using the Cyclix-DFT feature \cite{sharma2021real} in the SPARC code \cite{xu2021sparc, ghosh2017sparc1, ghosh2017sparc2}.}
\label{fig:illustration}
\end{figure}

We perform Kohn-Sham DFT simulations to calculate the elastic properties of the nanotubes using the Cyclix-DFT \cite{sharma2021real} feature in the state-of-the-art real-space code SPARC \cite{xu2021sparc, ghosh2017sparc1, ghosh2017sparc2}. In particular, Cyclix-DFT can exploit the cyclic and helical symmetry in the system to reduce all computations to the fundamental domain \cite{sharma2021real, ghosh2019symmetry, banerjee2016cyclic}, which in the current context is a unit cell with only $3$ atoms: 1 metal atom and 1 chalcogen atom of each type,  a situation that is true even on the application of axial and/or torsional deformations (Fig.~\ref{fig:illustration}). This symmetry-adaption provides tremendous reduction in the computational expense, with many of the simulations performed here impractical for even state-of-the-art DFT codes, even on powerful supercomputers \cite{banerjee2018two, xu2021sparc}. For instance, a MoSSe  nanotube of diameter $\sim 8.5$ nm subject to a twist of $6\times10\textsuperscript{-4}$ rad/Bohr has $219,888$ atoms in the unit cell when traditional periodic boundary conditions are prescribed, a system that is clearly intractable using standard approaches. Note that the accuracy of Cyclix-DFT  has been verified by not only comparison with  established DFT codes \cite{sharma2021real}, but also through its ability to make accurate predictions in a number of physical applications \cite{codony2021transversal, kumar2021flexoelectricity, kumar2020bending, 10.1088/1361-6528/ac1a90, bhardwaj2021torsional}.

In all the simulations, we employ pseudopotentials from the SPMS collection \cite{SPMS}, which is a set of transferable and soft optimized norm-conserving Vanderbilt (ONCV) pseudopotentials \cite{hamann2013optimized} with nonlinear core correction (NLCC). The accuracy of the pseudopotentials in the current context is confirmed by the very good agreement of the computed  equilibrium geometry (Supplementary Material) with previous DFT results for Janus TMD nanotubes \cite{luo2019electronic, mikkelsen2021band, tang2018janus, wang2018mechanical, bolle2021structural} as well as monolayers \cite{haastrup2018computational, bolle2021structural, shi2018mechanical, zhao2015ultra}. We employ the semilocal Perdew–Burke–Ernzerhof (PBE)  exchange-correlation functional \cite{perdew1996generalized}, which is considered to accurately describe properties/behavior for TMD systems \cite{haastrup2018computational, zhou20192dmatpedia, wang2016strain, xiao2014theoretical, li2014strain,ataca2012stable, chang2013orbital, amin2014strain, guo2014tuning, luo2019electronic, mikkelsen2021band, tang2018janus, wang2018mechanical, bolle2021structural, shi2018mechanical, zhao2015ultra}, as validated by experimental measurements \cite{chen2003titanium, nath2003superconducting, nath2002nanotubes, nath2001mose2, klots2014probing, ugeda2014giant, hill2016band, novoselov2005two, coleman2011two, lu2017janus, zhang2017janus}. Indeed, the use of more advanced density functionals such as hybrids and/or inclusion of spin orbit coupling (SOC) are not expected to  change the elastic properties noticeably, given that small strains are accompanied by small perturbations of electron density with respect to the undeformed nanotube, which translates to significant cancellations of error while taking energy differences.   This is expected to be particularly true in the current context, given that the difference between PBE and hybrid functionals for calculating ground state electron density has been found to be relatively small for TMD monolayers, both with and without  SOC \cite{kumar2020bending}.

We calculate the Young's modulus $E$, Poisson's ratio $\nu$, and torsional modulus $K$ by fitting the data to the relations:
\begin{eqnarray}
\mathcal{E}(0, \varepsilon, \tilde{\varepsilon}^*(\varepsilon)) \equiv \min_{\tilde{\varepsilon}} \mathcal{E}(0, \varepsilon, \tilde{\varepsilon}) = \mathcal{E} (0,0,0) + \frac{1}{2} E \varepsilon^2 \,, \\
\tilde{\varepsilon}^* = - \nu \varepsilon \,, \\
\mathcal{E}(\theta, \varepsilon^*(\theta), 0)  \equiv \min_{\varepsilon} \mathcal{E}(\theta, \varepsilon, 0)  =  \mathcal{E} (0,0,0) + \frac{1}{2} K \theta^2 \,, 
\end{eqnarray}
where $\mathcal{E}(\theta, \varepsilon, \tilde{\varepsilon})$ is the  energy density --- value at the electronic ground state  corresponding to the force-relaxed atomic configuration --- for twist density $\theta$, axial strain $\varepsilon$, and circumferential/hoop strain $\tilde{\varepsilon}$, with both energy and twist densities defined to be  per unit length of the nanotube. The superscript $^*$ is used to denote the value of the quantity that minimizes the energy density. The numerical parameters in the Cyclix-DFT simulations, including real-space grid spacing, reciprocal space grid spacing for Brillouin zone integration, radial vacuum, and cell/atom structural relaxation tolerances are selected to ensure that the reported Young's and torsional moduli are converged to within 1\% of their values. This translates to the ground state energy being accurate to within $10^{-5}$ Ha/atom, which is required to capture the exceedingly small energy differences that exist for the mechanical deformations considered in this work, which  have been chosen to be small enough so as to have strains are in accordance with those used in experiments. \cite{levi2015nanotube, divon2017torsional, nagapriya2008torsional, kaplan2007mechanical, kaplan2006mechanical}.

\section{Results and discussion} \label{Sec:Results}
We have performed the aforedescribed symmetry-adapted Kohn-Sham DFT simulations to calculate the elastic properties of the twenty-seven select Janus TMD nanotubes, in both armchair and zigzag configurations, at their equilibrium diameters. The results so obtained have been summarized  in Table~\ref{tab:Properties_table} and Fig.~\ref{fig:violin}, which we discuss in detail below. All simulation data  can be found in the Supplementary Material.  Note that the results for the torsional moduli are reported in terms of the diameter-independent quantity referred to as the torsional modulus coefficient $\hat{k} = K/d^3$ \cite{bhardwaj2021torsional}, where $d$ is the  nanotube diameter.  Also note that both the  Young's and torsional moduli are reported in units of N/m rather than N/m$^2$, since the latter requires an assumption on the thickness of the nanotube, whose value is not clearly defined \cite{huang2006thickness}.

    
    	\begin{table}[h!]
 
		\centering

\caption{Young's modulus ($E$), Poisson's ratio ($\nu$), and torsional modulus coefficient ($\hat{k}$) for the twenty-seven select Janus TMD nanotubes from Kohn-Sham DFT calculations.}

\label{tab:Properties_table}
\renewcommand{\arraystretch}{0.74}
		\resizebox{0.7\textwidth}{!}{
			\begin{tabular}{|c|c|c|c|c|c|c|c|c|}
			\hline 		
\multirow{3}{*}{M}&\multicolumn{8}{c|}{MSSe}\\
\cline{2-9}
& \multicolumn{4}{c|}{Armchair}&\multicolumn{4}{c|}{Zigzag}\\
\cline{2-9}
&{D (nm)}  &{E (N/m)} & {$\nu$}&{$\hat{k}$ (N/m)} &{D (nm)} &{E (N/m)}   & {$\nu$} &{$\hat{k}$ (N/m)}\\
					\hline

{W} & {8.8$\pm$ 0.5} &{106}& {0.25} &{39}  &{9.0 $\pm$ 0.5}  & {114} & {0.19}& {38}  \\ 
{Mo}&{8.4 $\pm$ 0.2 }  &{96} & {0.28}&{33}&{8.3 $\pm$ 0.1} &  {104} &{0.22} & {33}\\
{Cr} & {7.2 $\pm$ 0.3}& {87} & {0.35}& {30}& {7.5 $\pm$ 0.4}&   {97}& {0.25} & {29}\\
{Ta} &{ 11.1 $\pm$ 1.5 }& {80}& {0.41}& {25}&{11.2 $\pm$ 1.5}& {90} & {0.29}& {25}\\ 	
{V} & {14.6 $\pm$ 2 }&  {72}& {0.32} &{23}& { 14.2 $\pm$ 2.3}& {79}& {0.24}& {27}\\ 
{Nb}& { 13.1 $\pm$ 1.4 }&  {68}& {0.35}& {23}& { 14.1 $\pm$ 1.4 } &{76}& {0.26}& {23}\\ 
{Hf}& {9.5 $\pm$ 0.5 }&  {62}& {0.20}&{22}& {10.4 $\pm$ 0.5 } &  {65}& {0.15}& {22}\\ 
{Zr} & {15.5 $\pm$ 3 } & {57} & {0.19}& {20} & {14.9 $\pm$ 2.4} &  {59}  & {0.14}& {20}\\
{Ti} &{ 43.9 $\pm$ 5.9}&{60} & {0.21}&  {19} & { 44.9 $\pm$ 4.9}& {62} & { 0.20}& {20}  \\ 

\hline
\multirow{3}{*}{M}&\multicolumn{8}{c|}{MSTe}\\
\cline{2-9}
& \multicolumn{4}{c|}{Armchair}&\multicolumn{4}{c|}{Zigzag}\\
\cline{2-9}
&{D (nm)}  &{E (N/m)} & {$\nu$}&{$\hat{k}$ (N/m)} &{D (nm)} &{E (N/m)}   & {$\nu$} &{$\hat{k}$ (N/m)}\\
					\hline
			
{W} & {3.8 $\pm$ 0.1 }  & {80}& {0.42}& {32} & {3.8 $\pm$ 0.1 }  & {99} & {0.31} & {35}\\ 
{Mo}&{3.8 $\pm$ 0.1 }  & {74} & {0.48}& {28} & {3.8 $\pm$ 0.1 } & {93} & {0.40} & {30}\\ 
{Cr} & {3.1 $\pm$ 0.1 }&{60}& {0.54}  & {22}  & {3.1 $\pm$ 0.1 } & {83}& {0.36} & {25} \\
{Ta} &{4.0 $\pm$ 0.1}  & {65} & {0.53}& {24} &{4.1 $\pm$ 0.06}& {80}  & {0.42} &  {28} \\
{V} & {4.4 } & {48}& {0.43}&  {21} & {4.4 $\pm$ 0.06 } &  {65}& {0.47} &  {21}\\
{Nb}& {4.4 $\pm$ 0.1}  & {47}& {0.55}& {23} & {4.4 $\pm$ 0.1 } & {69}& {0.39} &  {20} \\
{Hf}& {4.4 $\pm$ 0.4 } &{43} & {0.30}&  {19} & {4.4 $\pm$ 0.4 } & {51}&  {0.24} & {18} \\
{Zr} & { 6.1 $\pm$ 0.2}  & {39}& {0.29}& {16} & {6.1 $\pm$ 0.2 }& {45}& {0.20}&  {15} \\
{Ti} &{10.0 $\pm$ 0.4 }&{46} & {0.34}&  {16}&{9.7 $\pm$ 0.6 }&   {48} & {0.34} & {14}\\

\hline			
\multirow{3}{*}{M}&\multicolumn{8}{c|}{MSeTe}\\
\cline{2-9}
& \multicolumn{4}{c|}{Armchair}&\multicolumn{4}{c|}{Zigzag}\\
\cline{2-9}
&{D (nm)}  &{E (N/m)} & {$\nu$}&{$\hat{k}$ (N/m)} &{D (nm)} &{E (N/m)}   & {$\nu$} &{$\hat{k}$ (N/m)}\\
					\hline
			
{W} & {6.7 $\pm$ 0.1 }  & {77} &{0.26} &{31}& {7.1 $\pm$ 0.1 }& {85} & {0.18} & {31}\\
{Mo}&{6.6 $\pm$ 0.3 }& {69}& {0.32}  & {27}& {6.5 $\pm$ 0.3 } & {81}& {0.19} &  {27}\\
{Cr} & {5.6 $\pm$ 0.2 }&{62} & {0.44} & {22} & {5.6 $\pm$ 0.1 } & {73} & {0.34} & {22}\\
{Ta} &{7.6 $\pm$ 0.3 }& {58} & {0.40}& {20}& { 7.6 $\pm$ 0.2}&{66} & {0.31} &  {20}\\
{V} & {8.4 $\pm$ 0.6 }&  {52} & {0.38}&{17}& { 8.2 $\pm$ 0.5}&  {59}& {0.27} & {25}\\
{Nb}& { 8.7 $\pm$ 0.8}& {53} & {0.38} &  {18}& { 8.9 $\pm$ 0.8}& {59}& {0.28} & {16}\\
{Hf}& {8.2 $\pm$ 0.5 } & {40}& {0.15} & {17}& {8.1 $\pm$ 0.6 } &{44} & {0.10} & {16}\\
{Zr} & {16.6 $\pm$ 3 } & {38} & {0.20}&{14} & {15.8 $\pm$ 2.2 } &{42}& {0.12} & {14}\\
{Ti} &{39.3 $\pm$ 10.9 } & {41} & {0.16}& {13}& { 28.7 $\pm$ 4.9}& {40}& {0.15} & {13}\\  

            \hline

        	\end{tabular}}

	\end{table}

\begin{figure}[ht!]
        \centering
        \includegraphics[width=0.85\textwidth]{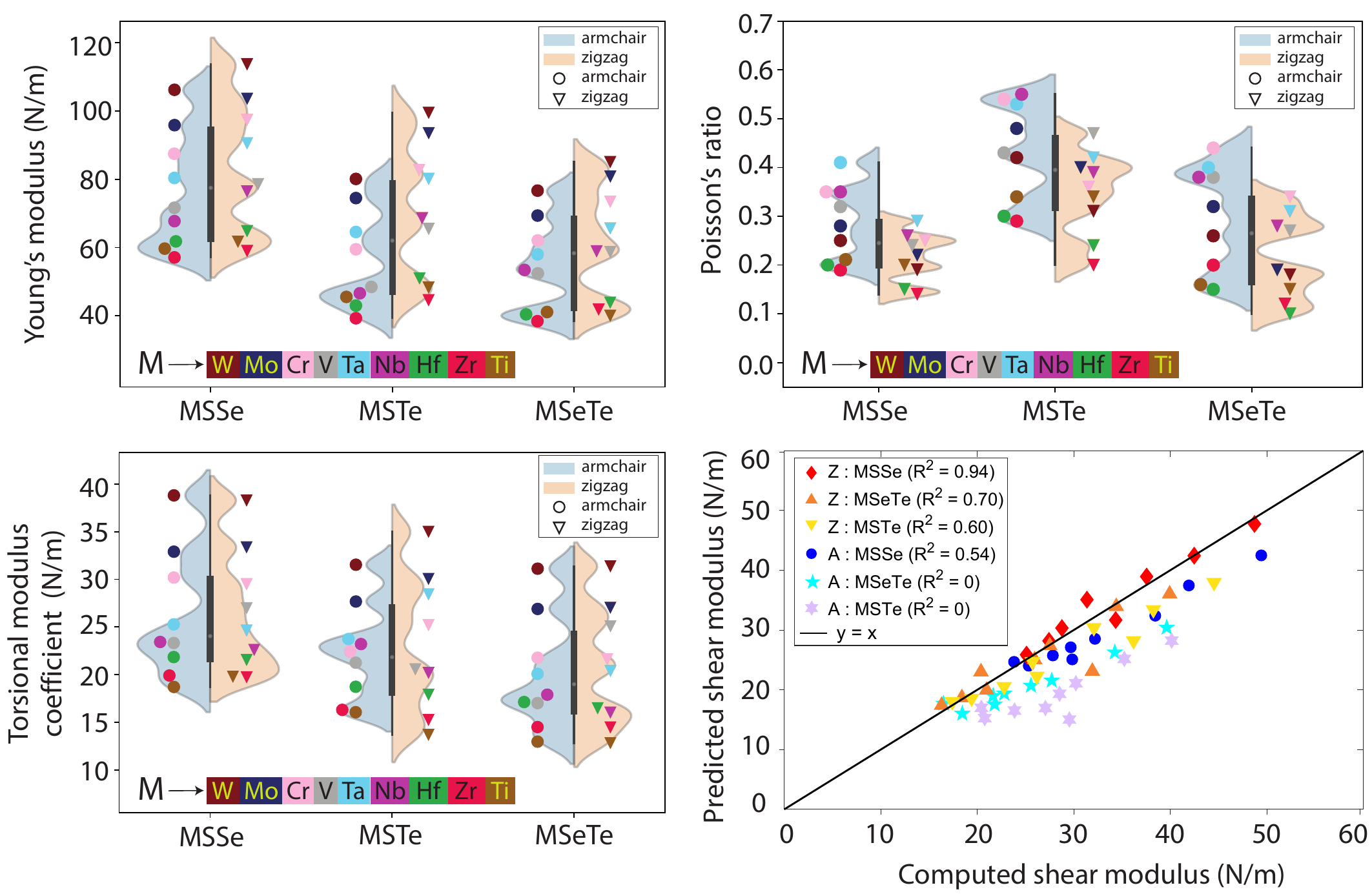}
        \caption{Young's modulus ($E$), Poisson's ratio ($\nu$), and torsional modulus coefficient ($\hat{k}$) for the twenty-seven select armchair and zigzag Janus TMD nanotubes. Also shown is the shear modulus computed from the torsional modulus coefficient vs. that predicted from the Young's modulus and Poisson's ratio. The $R^2$ values listed in the legend represent the linear regression's coefficient of determination.}
      \label{fig:violin}
    \end{figure}

The equilibrium radii generally follow the trend: MSTe $<$ MSeTe $<$ MSSe, which can be correlated to the difference in electronegativity between the chalcogens, i.e., larger  electronegativity differences result in  smaller  equilibrium diameters, an observation that is in agreement with previous DFT results for M$=$\{Nb, Ta, Mo, W\} and X,Y$=$\{S, Se, Te\} \cite{zhao2015ultra}.  The computed equilibrium diameters for these systems are also in excellent agreement with Ref~\cite{zhao2015ultra}, the maximum difference being 0.4 nm, which occurs for NbSSe. In terms of comparison with Ref.~\cite{bolle2021structural}, which also employs DFT to predict  the equilibrium diameters for all the materials studied here, while there is good agreement for nanotubes with smaller diameters, the difference increases with nanotube diameter, reaching a maximum of 22.8 nm for TiSeTe. This is a consequence of Ref.~\cite{bolle2021structural}  using extrapolation from the data for small diameters --- the current work employs interpolation, with data points on either side of the equilibrium diameter--- whereby larger errors are accumulated when the equilibrium diameter is farther away from the region where Kohn-Sham calculations have actually been performed.

We observe from the results that the Young's moduli and torsional modulus coefficients for the Janus TMD nanotubes lie between the corresponding values for the parent TMD nanotubes \cite{bhardwaj2021torsional}. In addition, we find that they follow the trend: MSSe $>$ MSTe $>$ MSeTe, which is the  reverse of that for the metal-chalcogen bond lengths. Indeed, increased overlap of the orbitals at shorter interatomic distances is expected to result in stronger bonds. In terms of individual values, WSSe and ZrSeTe have the largest and smallest Young's moduli, respectively, while WSSe and TiSeTe have the largest and smallest torsional modulus coefficients, respectively. Notably, carbon nanotube's moduli ($E=345$ N/m \cite{treacy1996exceptionally} and $\hat{k} = 117$ N/m \cite{sharma2021real}) are close to a factor of three higher than the largest values here, a likely consequence of the very strong covalent carbon-carbon bonds. Regarding the Poisson's ratio, we find that the armchair CrSTe, TaSTe, NbSTe nanotubes have values larger than 0.5 --- theoretical limit for isotropic materials \cite{ting2005poisson} --- which suggests their anisotropic nature, a result that we further confirm below. Note that we are not aware of any theoretical or experimental results in literature against which we can compare the values reported in this work. Indeed, the predictions made for MoSSe in Ref.~\cite{wang2018mechanical} cannot be compared, given the significantly smaller  diameter chosen there compared to the equilibrium value used here. 

We also observe from the results that  anisotropy with respect to armchair and zigzag configurations follows the ordering:  MSTe $>$ MSeTe $>$ MSSe, which can likely be attributed to the level of dissimilarity between the chalcogens. To confirm this, we compute the effective shear modulus using the relation $G = 4 \hat{k}/\pi$ --- corresponds to the nanotube being modeled as a homogeneous circular tube made of isotropic material, subject to an external twist --- and compare it against that predicted by the  relation featuring the Young's modulus and Poisson's ratio for isotropic materials: $G = E/2(1+\nu)$.  The results so obtained are summarized in Fig.~\ref{fig:violin}, from which it is clear that the disagreement between the computed and predicted shear moduli follows the same trend as that stated above for the difference in values between armchair and zigzag configurations,  indicative of the relative degree of anisotropy between the different groups. In addition, the armchair nanotubes are significantly more anisotropic compared to their zigzag counterparts.

  \begin{figure}[ht!]
        \centering
        \includegraphics[width=0.85\textwidth]{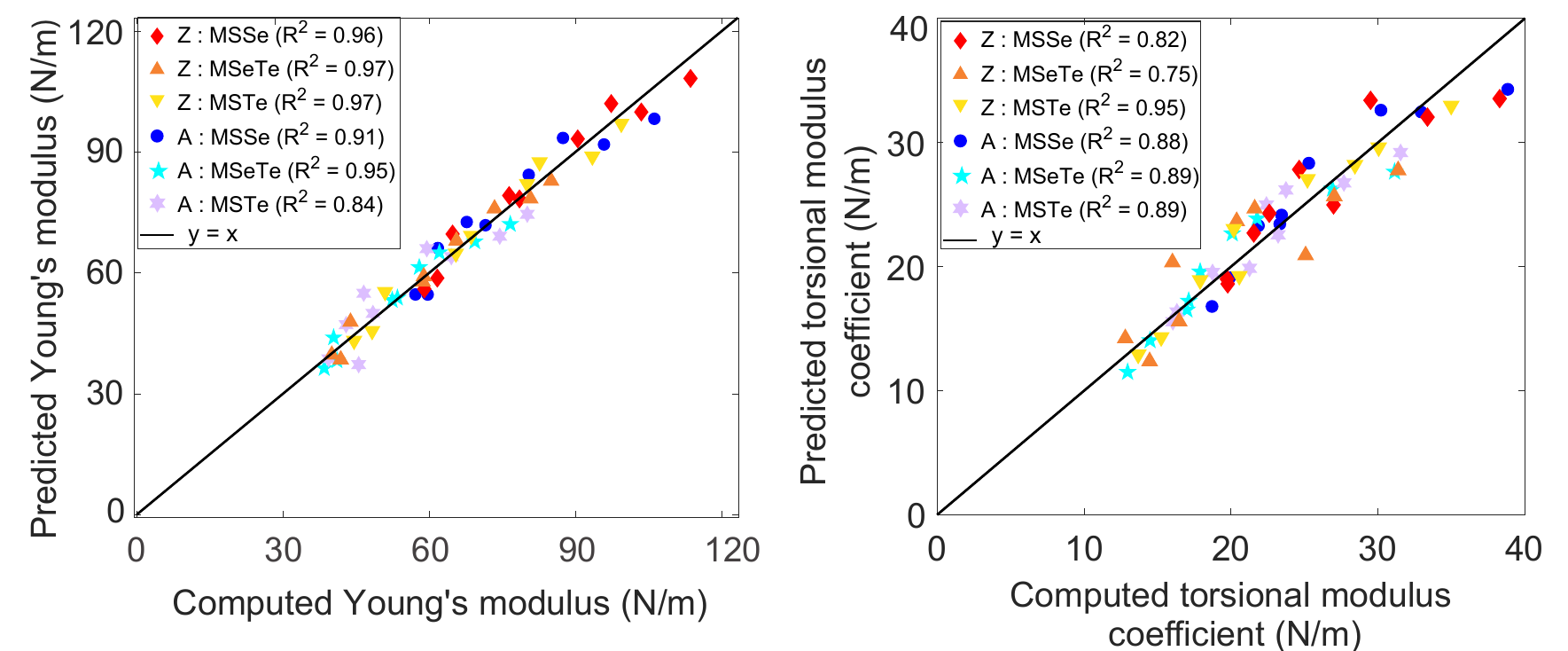}
        \caption{Computed Young's modulus and torsional modulus coefficient vs. that predicted by the model based on linear regression. The regression model has the following features: metal-chalcogen bond lengths, difference in electronegativity between the chalcogen atoms, and sum of ionization potential and electron affinity of metal and chalcogens respectively. The $R^2$ values listed in the legend represent the linear regression's coefficient of determination.}
       \label{fig:Regression}
    \end{figure} 
 
The aforedescribed results suggest that the nature/strength of the metal-chalcogen bonds as well as the level of dissimilarity between the chalcogen atoms plays a significant role in determining the Young's and torsional  moduli of Janus TMD nanotubes. In view of this, we develop a regression model with the following features: metal-chalcogen bond lengths, difference in electronegativity between the chalcogens, and sum of the ionization potential and electron affinity for the metal and chalcogens, respectively. In particular, we carry out a linear regression on the computed Young's moduli and torsional modulus coefficients, whose results are presented in Fig.~\ref{fig:Regression}. The fit is reasonably good, which confirms that the features chosen for our model do play a notable role in deciding the elastic properties for Janus TMD nanotubes. It is worth noting that we can further improve the quality of the fit by using a higher-order regression. However, this strategy comes with the possibility of overfitting, and therefore not adopted here.


\section{Concluding remarks} \label{Sec:Conclusions}
In this work, we have calculated the elastic properties of  select single-walled Janus TMD nanotubes from first principles Kohn-Sham DFT. In particular, considering the twenty-seven Janus TMD nanotubes that have previously been predicted to be stable,  we have performed cyclic and helical symmetry-adapted electronic structure simulations to compute the Young's moduli, Poisson's ratios, and torsional moduli for the armchair and zigzag versions of these materials at their equilibrium diameters. We have found the following trend in the moduli: MSSe $>$ MSTe $>$ MSeTe, while their anisotropy with respect to armchair and zigzag configurations has the ordering:  MSTe $>$ MSeTe $>$ MSSe. We have confirmed this anisotropy and ordering between the different groups by computing the shear modulus from the torsional modulus using an isotropic elastic continuum model, and comparing it with the value predicted from the isotropic relation featuring Young's modulus and Poisson's ratio. We have also developed a reasonably accurate  model for Young's and torsional  moduli of Janus TMD nanotubes based on linear regression, with the following features: metal-chalcogen bonds' nature/characteristics and the difference in electronegativity between the chalcogens. 

The electronic response of Janus TMD nanotubes to mechanical deformations presents itself as an interesting research problem worthy of pursuit, having a number of potential applications in semiconductor devices.  In addition, given their anisotropic nature, the effect of chirality on the electromechanical response of Janus TMD nanotubes is also a worthy subject for future research.


\section*{Acknowledgements} 
The authors gratefully acknowledge the support of the US National Science Foundation (CAREER-1553212 and MRI-1828187).

\end{document}